\documentclass[aps,pre,twocolumn,showpacs,amsmath,amssymb,floatfix]{revtex4-1}
\usepackage[utf8x]{inputenc} 
\usepackage{graphicx}
\usepackage{multirow}
\usepackage{rotating}

\begin{document}

\title{Wang-Landau sampling: a criterion for halting the simulations}

\author{A. A. Caparica}

\affiliation{Instituto de F\'{\i}sica, Universidade Federal de
Goi\'{a}s.  C.P. 131, CEP 74001-970, Goi\^{a}nia, GO, Brazil \\
}

\begin{abstract}
In this work we propose a criterion to finish the simulations of the Wang-Landau sampling. Instead of determining a final modification factor for all simulations and every sample sizes, we investigate the behavior of the temperature of the peak of the specific heat during the simulations and finish them when this value varies bellow a given limit. As a result, different runs stop at different final modification factors. We show that in place of the temperature of the peak of the specific heat one can adopt alternatively the integrated heat transfer as a reference quantity. We apply this technique to the two-dimensional Ising model and a homopolymer. We verify that for the Ising model the mean order of the final modification factors are roughly the same for all lattice sizes, but for the homopolymer the order of the final modification factors increases with increasing polymer sizes. The results show that the simulations can be halted much earlier then it is conventional in Wang-Landau sampling, but manifold finite-size simulations are required in order to obtain accurate results. A brief application to the three-dimensional Ising model is also available.
\end{abstract}

\maketitle

\section{Introduction}
\vspace{.25cm}
Wang-Landau sampling (WLS)\cite{wls1,wls2} can be considered a well-established Monte Carlo method,
since it has been applied efficiently to many systems. Nevertheless the method is still in
development and new ideas have contributed to increase the degree of efficiency and accuracy of the
algorithm.

The method is based on the fact that if one performs a random walk in energy space with a probability
proportional to the reciprocal of the density of states, a flat histogram is generated for the energy
distribution. Since the density of states produces huge numbers, instead of estimating $g(E)$, the
simulation is performed for $S(E)\equiv\ln g(E)$. At the beginning we set $S(E)=0$ for all energy levels.
The random walk in the energy space runs through all energy levels from $E_{min}$ to $E_{max}$ with a
probability $p(E\rightarrow E')=\min(\exp{[S(E)-S(E')]},1)$, where $E$ and $E'$ are the energies of the
current and the new possible configurations. Whenever a configuration is accepted we update $H(E')\rightarrow
H(E')+1$ and $S(E')\rightarrow S(E')+F_{i}$, where $F_{i}=\ln f_{i}$, $f_{0}\equiv e=2.71828...$ and
$f_{i+1}=\sqrt{f_{i}}$ ($f_{i}$ is the so-called modification factor). The flatness of the histogram is checked
after a number of Monte Carlo steps and usually the histogram is considered flat if $H(E)>0.8\langle H \rangle$,
for all energies, where $\langle H \rangle$ is an average over energies. If the flatness condition is fulfilled we
update the modification factor to a finer one and reset the histogram $H(E)=0$. The original version of WLS
prescribes that simulations should be in general halted when $f\sim1+10^{-8}$. Having in hand the density of states,
one can calculate the canonical average of any thermodynamic variable $X$ as

\begin{equation}\label{mean}
\langle X\rangle_T=\dfrac{\sum_E \langle X\rangle_E g(E) e^{-\beta E}}{\sum_E g(E) e^{-\beta E}} ,
\end{equation}
where $\langle X\rangle_E$ is the microcanonical average accumulated during the simulations and $\beta=1/k_BT$,
$k_B$ is the Boltzmann constant, and $T$ is the temperature.

Recent works \cite{caparica,caparica1,caparica2} have demonstrated that (a) instead of updating the density of states after every
move, one ought to update it after each Monte Carlo sweep (MCS) \cite{mcs}; (b) WLS should be carried out only up to
$\ln f=\ln f_{final}$ defined by the canonical averages during the simulations; and (c) the microcanonical
averages should not be accumulated before $\ln f \leq ln f_{micro}$ defined by the microcanonical averages
during the simulation. The adoption of these easily implementable changes leads to more accurate results
and saves computational time. They investigated the behavior of the maxima of the specific heat
\begin{equation}\label{cv}
 C(T)=\langle(E-\langle E\rangle)^2\rangle/T^2
\end{equation}
and the susceptibility
\begin{equation}\label{ki}
 \chi(T)=L^2\langle(m-\langle m\rangle)^2\rangle/T,
\end{equation}
where $E$ is the energy of the configurations and $m$ is the corresponding magnetization per spin, during the
WLS for the Ising model on a square lattice. They observed (as in \cite{belardinelli,belardinelli2,belardinelli3,swetnam})
that a considerable part of the conventional Wang-Landau simulation is not very useful because the error saturates.
They demonstrated in detail that in general no single simulation converges to the true value, but to a particular value of a Gaussian distribution of results around the correct value. The saturation of the error coincides with the convergence to this value.
Continuing the simulations beyond this limit leads to irrelevant variations in the canonical averages of all thermodynamic variables.
In order to define $f_{final}$ to a given model one should take a representative size ($L=32$ for the 2D Ising model
and $N=50$ for the homopolymer) and find out when the corresponding canonical averages obtained from a few independent
runs would come to steady values. In that study they concluded that $f_{final}$ should be $f_{13}$ and $f_{18}$ for
the 2D Ising model and the homopolymer, respectively.

In the present work we propose a criterion for finishing the simulations which turns the choice of $f_{final}$ automatic for each
independent run. As a result the simulated data become more accurate and one has no need to find out $f_{final}$ before
initiating the simulations. We found out also that two independent finite-size scaling procedures can lead to results
that do not agree within the errorbars and therefore the final results should be obtained from manifold independent procedures.

\maketitle

\section{A criterion for finishing the simulations}
\vspace{.25cm}
From time to time during the WLS the random walker pauses the simulation in order to check the
histogram for flatness.

\begin{figure}[!ht]\centering
\begin{center}
 \includegraphics[width=.9\linewidth]{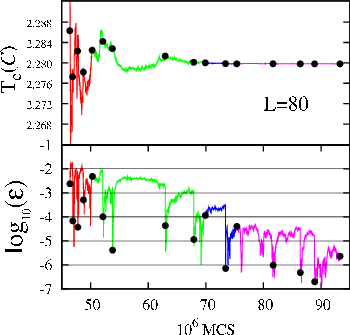}
\end{center}
\vspace{-.5cm}
\caption{(color online). Upper panel: Evolution of the temperature of the maximum of the specific heat during the WLS,
beginning from $f_{5}$ for a single run. The dots show where the modification factor was updated. Lower panel: Evolution
of the logarithm of the checking parameter $\varepsilon$ during the same simulation.}
\label{checking}
\end{figure}

Applying WLS to the two-dimensional Ising model, beginning from $f_{5}$,
we calculate the temperature of the peak of the specific heat defined in Eq.\eqref{cv} using the current $g(E)$ and
from this time on this mean value is updated whenever the histogram is checked for flatness.
When the histogram is considered flat, we save the value of the temperature of the peak of the specific heat $T_c(0)$.
We then update the modification factor $f_{i+1}=\sqrt{f_{i}}$ and reset the histogram $H(E)=0$.
During the simulations with this new modification factor
we continue calculating the temperature of the peak of the specific heat $T_c(t)$ whenever we check the histogram for
flatness and we also calculate the checking parameter

\begin{equation}\label{eps}
 \varepsilon=|T_c(t)-T_c(0)|.
\end{equation}

If the number of MCS before verifying the histogram for flatness is chosen not too large, say 10,000, then during the simulations with the same modification factor the checking parameter $\varepsilon$ is calculated many times.
The idea of the proposed criterion is that if $\varepsilon$ remains less than a predefined threshold limiting value, which we will
refer to as $limit$, until the histogram meets the flatness criterion for this WL level, then we save the density of
states and the microcanonical averages and stop the simulations. The expedient of observing the behavior of the specific heat to end simulations have been used in a more informal way in Ref. \cite{eisenbach}.

Top of Fig. \ref{checking} shows the evolution of the temperature of the peak of the specific heat beginning from $f_{5}$ calculated for $L=80$ as a function of the MCS using the $80\%$ flatness criterion. The dots show where the modification factor was updated. The lower panel shows the evolution of the
logarithm of the checking parameter $\varepsilon$ during the same simulation. In this
example one can see that for $limit=10^{-2}$, $10^{-3}$, $10^{-4}$ and $10^{-5}$, the simulation would be finished at $f_{8}$, $f_{13}$, $f_{15}$, and $f_{19}$, respectively.

The convergence of the temperature of the peak of the specific heat to a value of a Gaussian distribution around the true value
is assured by the convergence of the density of states itself. Zhou and Bhatt \cite{zhou_bhatt} demonstrated that when $f$ is
close to $1$ the relative error $\delta g/g=\delta\ln g$ scales as $\sqrt{\ln f}$.

\begin{figure}[!h]\centering
\begin{center}
 \includegraphics[width=.9\linewidth]{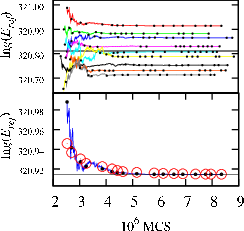}
\end{center}
\vspace{-.5cm}
\caption{(color online). Upper panel: Evolution of $\ln g(E_{ref})$ during the WLS,
beginning from $f_{11}$ for nine independent runs. The dots show where the modification factor was updated and the straight
line indicates the exact value $\ln g_{exact}(E_{ref})$ from Ref. \cite{beale} . Lower panel: The uppermost curve alone. The straight
line indicates the value $\ln g_{final}$ to which this simulation converges and the open circles correspond
to $\ln g_{final}+\sqrt{\ln f}$}
\label{zhou}
\end{figure}

In order to illustrate this convergence, using
the exact density of states provided by Beale \cite{beale} we calculated the exact temperature of the maximum of the specific
heat of a $32\times32$ lattice: $T_c=2.29392897$. For this temperature the exponential $e^{-E/k_BT}$ is maximum at $E_{ref}=-1424$.
We take this level as reference. In top of Fig. \ref{zhou} we show the evolution of $\ln g(E_{ref})$ during the simulations, beginning from $f_{11}$ for nine independent runs. One can see that they all do converge to steady, but different values.
In the lower panel we show only the uppermost one which converges to $\ln g_{final}=320.914628304468$, while the exact value
is $\ln g_{exact}(E_{ref})=320.810960956218$ \cite{beale}. The open cycles display $\ln g_{final}+\sqrt{\ln f}$. One can see that these results corroborate that the relative error do scales with $\sqrt{\ln f}$, but not with respect to the true value. Based on the results of Ref. \cite{caparica} we can presume that these values also fall into a Gaussian distribution around the exact value.

In order to define which would be the optimal value for the quantity $limit$, we performed
finite-size scaling simulations taking outcomes when the criterion mentioned above was satisfied for
$limit=10^{-3}$, $10^{-4}$, and $10^{-5}$.

According to finite-size scaling theory \cite{fisher1,fisher2,barber} from the definition of the
free energy one can obtain the zero field scaling expressions for the magnetization and the susceptibility, respectively by

\begin{equation}\label{exp1}
 m\approx L^{-\beta/\nu}\mathcal{M}(tL^{1/\nu}),
\end{equation}

\begin{equation}\label{exp2}
\chi \approx L^{\gamma/\nu}\mathcal{X}(tL^{1/\nu}).
\end{equation}

\noindent We see that the locations of the maxima of these functions scale asymptotically as
\begin{equation}\label{tc}
 T_c(L) \approx T_c+a_qL^{-1/\nu},
\end{equation}
where $a_q$ is a quantity-dependent constant, allowing then the determination of $T_c$.

Using these scaling functions and assuming $\nu=1$, we estimated the critical temperature and the critical
exponents $\beta$ and $\gamma$.

In our initial attempts of using this checking parameter we produced also outcomes when the simulations reached
the end of the modification factor $f_{13}$, as prescribed in Ref.\cite{caparica}, in order to reproduce those
results. Surprisingly we found out that two independent similar finite-size scaling procedures can lead to
very different results for the critical temperature, as $T_c=2.26864(24)$ and $T_c=2.26987(28)$, for example, which do not
agree within the errorbars.  In order to circumvent this difficult,
we performed ten independent Wang-Landau simulations for $L=32$, $36$, $40$, $44$, $48$, $52$, $56$, $64$, $72$ and $80$ with
$N=24$, $24$, $20$, $20$, $20$, $16$, $16$, $16$, $12$ and $12$ independent runs for each size, respectively.

\begin{table}[!h]
  \centering
  \begin{tabular}{c c c c }
  \hline\hline
      Exact & &$T_c$=2.2691853...  &  \\
  \hline
    $f_{13}$ & \hspace{.3cm} $\varepsilon<10^{-3}$ & \hspace{.3cm} $\varepsilon<10^{-4}$ & \hspace{.3cm} $\varepsilon<10^{-5}$ \\ 
  \hline
 2.26864(24) & \hspace{.3cm} 2.26845(21) & \hspace{.3cm} 2.26883(26) \hspace{.3cm} & 2.26889(25) \\
 2.26857(26) & \hspace{.3cm} 2.26870(31) & \hspace{.3cm} 2.26862(24) \hspace{.3cm} & 2.26869(20) \\
 2.26905(22) & \hspace{.3cm} 2.26905(26) & \hspace{.3cm} 2.26921(20) \hspace{.3cm} & 2.26920(18) \\
 2.26915(13) & \hspace{.3cm} 2.26941(17) & \hspace{.3cm} 2.26922(12) \hspace{.3cm} & 2.26918(11) \\
 2.26943(13) & \hspace{.3cm} 2.26955(19) & \hspace{.3cm} 2.26937(11) \hspace{.3cm} & 2.26937(11) \\
 2.26920(30) & \hspace{.3cm} 2.26939(32) & \hspace{.3cm} 2.26923(26) \hspace{.3cm} & 2.26903(26) \\
 2.26999(28) & \hspace{.3cm} 2.27000(27) & \hspace{.3cm} 2.26948(24) \hspace{.3cm} & 2.26938(20) \\
 2.26871(28) & \hspace{.3cm} 2.26877(34) & \hspace{.3cm} 2.26894(22) \hspace{.3cm} & 2.26901(20) \\
 2.26912(24) & \hspace{.3cm} 2.26889(32) & \hspace{.3cm} 2.26899(22) \hspace{.3cm} & 2.26878(15) \\
 2.26987(28) & \hspace{.3cm} 2.27023(34) & \hspace{.3cm} 2.26964(27) \hspace{.3cm} & 2.26948(23) \\
\hline
 2.26917(15) & \hspace{.3cm} 2.26925(18) & \hspace{.3cm} 2.26916(10) \hspace{.3cm} & 2.26910(10) \\
 \hline\hline
  \end{tabular}
  \caption{Ten finite-size results for the critical temperature for outputs at the end
  of $f_{13}$ and for decreasing threshold $limit$, using the $80\%$ flatness criterion. The
  last line shows the average values over all the runs.}
  \label{table1}
  \end{table}

\begin{table}[!h]
  \centering
  \begin{tabular}{c c c c }
  \hline\hline
      Exact & \hspace{.3cm} $\gamma$=1.75        \\
  \hline
    $f_{13}$ & \hspace{.3cm} $\varepsilon<10^{-3}$ & \hspace{.3cm} $\varepsilon<10^{-4}$ & \hspace{.3cm} $\varepsilon<10^{-5}$ \\ [0.5ex]
  \hline
 1.7618(41) & \hspace{.3cm}  1.7615(34) & \hspace{.3cm} 1.7615(45) & \hspace{.3cm} 1.7586(34) \\
 1.7584(69) & \hspace{.3cm}  1.7579(86) & \hspace{.3cm} 1.7599(47) & \hspace{.3cm} 1.7573(36) \\
 1.7642(58) & \hspace{.3cm}  1.7614(69) & \hspace{.3cm} 1.7585(53) & \hspace{.3cm} 1.7580(38) \\
 1.7550(42) & \hspace{.3cm}  1.7534(50) & \hspace{.3cm} 1.7569(41) & \hspace{.3cm} 1.7579(23) \\
 1.7643(29) & \hspace{.3cm}  1.7629(46) & \hspace{.3cm} 1.7624(22) & \hspace{.3cm} 1.7656(32) \\
 1.7608(46) & \hspace{.3cm}  1.7606(44) & \hspace{.3cm} 1.7606(34) & \hspace{.3cm} 1.7642(34) \\
 1.7546(40) & \hspace{.3cm}  1.7560(33) & \hspace{.3cm} 1.7578(31) & \hspace{.3cm} 1.7584(21) \\
 1.7663(70) & \hspace{.3cm}  1.7661(76) & \hspace{.3cm} 1.7614(55) & \hspace{.3cm} 1.7603(41) \\
 1.7601(48) & \hspace{.3cm}  1.7583(58) & \hspace{.3cm} 1.7616(53) & \hspace{.3cm} 1.7601(31) \\
 1.7632(51) & \hspace{.3cm}  1.7623(51) & \hspace{.3cm} 1.7648(41) & \hspace{.3cm} 1.7649(26) \\
 \hline
 1.7609(12) & \hspace{.3cm}  1.7600(11) & \hspace{.3cm} 1.7605(10) & \hspace{.3cm} 1.7605(10) \\
 \hline\hline
  \end{tabular}
  \caption{Ten finite-size scaling results for the critical exponent $\gamma$ for outputs at the end
  of $f_{13}$ and for decreasing threshold $limit$, using the $80\%$ flatness criterion. The
  last line shows the average values over all the runs.}
  \label{table2}
  \end{table}

In Table \ref{table1} we show the results for the critical temperature obtained from these
ten independent finite-size scaling simulations and the mean values in the last line.  Each
temperature in this table is the mean value between the extrapolations of the temperatures
of the peaks of the specific heat and the susceptibility. One can see that each single
result can be particularly bad or good, but the average values for the temperature are excellent even for the less stringent $limit=10^{-3}$.

This criterion for finishing the WLS has two main advantages. First of all it is not necessary to determine $f_{final}$ for
a representative size, as prescribed in Ref.\cite{caparica} since it is defined automatically for each independent run.
The second convenience is that different runs can proceed up to different final modification factors, depending on the
evolution of the simulation.
\vspace{.3cm}
\begin{table}[!h]
  \centering
  \begin{tabular}{c c c c }
  \hline\hline
      Exact & \hspace{.3cm} $\beta$=0.125      \\
  \hline
    $f_{13}$ & \hspace{.2cm} $\varepsilon<10^{-3}$ & \hspace{.2cm} $\varepsilon<10^{-4}$ & \hspace{.2cm} $\varepsilon<10^{-5}$ \\ [0.5ex]
  \hline
 0.1263(21)  & \hspace{.2cm} 0.1276(17)  & \hspace{.2cm} 0.1255(22)  & \hspace{.2cm} 0.1251(20) \\
 0.1281(17)  & \hspace{.2cm} 0.1284(19)  & \hspace{.2cm} 0.1280(20)  & \hspace{.2cm} 0.1267(14) \\
 0.1263(17)  & \hspace{.2cm} 0.1268(19)  & \hspace{.2cm} 0.1252(17)  & \hspace{.2cm} 0.1245(15) \\
 0.1256(12)  & \hspace{.2cm} 0.1249(13)  & \hspace{.2cm} 0.1251(11)  & \hspace{.2cm} 0.1248(11) \\
 0.1255(15)  & \hspace{.2cm} 0.1260(15)  & \hspace{.2cm} 0.1256(14)  & \hspace{.2cm} 0.1254(13) \\
 0.1257(19)  & \hspace{.2cm} 0.1252(25)  & \hspace{.2cm} 0.1253(16)  & \hspace{.2cm} 0.1256(18) \\
 0.1216(19)  & \hspace{.2cm} 0.1211(16)  & \hspace{.2cm} 0.1235(16)  & \hspace{.2cm} 0.1235(14) \\
 0.1260(18)  & \hspace{.2cm} 0.1257(21)  & \hspace{.2cm} 0.1248(13)  & \hspace{.2cm} 0.1244(10) \\
 0.1250(19)  & \hspace{.2cm} 0.1264(25)  & \hspace{.2cm} 0.1257(17)  & \hspace{.2cm} 0.1258(11) \\
 0.1219(23)  & \hspace{.2cm} 0.1221(28)  & \hspace{.2cm} 0.1227(22)  & \hspace{.2cm} 0.1234(18) \\
 \hline
 0.12520(63) & \hspace{.2cm} 0.12541(72) & \hspace{.2cm} 0.12514(44) & \hspace{.2cm} 0.12491(32) \\
 \hline\hline
  \end{tabular}
  \caption{Ten finite-size scaling results for the critical exponent $\beta$ for outputs at the end
  of $f_{13}$ and for decreasing threshold $limit$, using the $80\%$ flatness criterion. The
  last line shows the average values over all the runs.}
  \label{table3}
  \end{table}

In Table \ref{table2} we show the finite-size scaling results for the exponent $\gamma$.
As in Ref.\cite{caparica}, they are a little bit above the exact value. For calculating
the exponent $\beta$ we used the averaged temperatures obtained in
Table \ref{table1} for each case. In Table \ref{table3} we present the results for $\beta$.
Again we can see that the results are in very close agreement with the exact value even for $limit=10^{-3}$.

\section{An alternative quantity for calculating the checking parameter}
\vspace{.25cm}

The use of the peak of the specific heat for calculating the checking parameter may in some cases be tricky since there are systems that exhibit more then one peak or it may be bad behaved. In such situations, another quantity may be used, namely, the heat transfer per unit calculated by the integrated specific heat in a broad temperature interval.

\begin{equation}\label{heat}
Q=\frac{1}{N}\int_{T_i}^{T_f} C(T) dT
\end{equation}
where $C(T)$ is the specific heat defined by Eq.\eqref{cv} and $N$ is the number of units in the system (number of spins,
monomers, etc.).

The checking parameter $\varepsilon$ can then be defined analogously as
\begin{equation}\label{eps_heat}
 \varepsilon=|Q(t)-Q(0)|.
\end{equation}
where as before, $Q(0)$ is the last value calculated in the previous modification factor and $Q(t)$ are the values calculated during
the current WL level. The integrals in Eq. \eqref{heat} were calculated for $T_i=1.0$ and $T_f=4.0$, using a step $\Delta T=0.01$.

\begin{figure}[!t]\centering
\begin{center}
 \includegraphics[width=.9\linewidth]{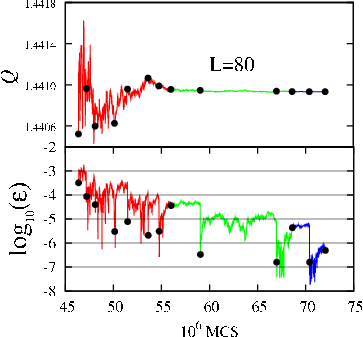}
\end{center}
\vspace{-.7cm}
\caption{(color online). Upper panel: Evolution of the temperature of the maximum of the specific heat during the WLS,
beginning from $f_{5}$ for a single run. The dots show where the modification factor was updated. Lower panel: Evolution
of the logarithm of the checking parameter $\varepsilon$ calculated using the quantity $Q$ during the same simulation.}
\label{checking_int}
\end{figure}

In Fig. \ref{checking_int} we show the evolution of the heat transfer per spin during the WLS and the logarithm of the checking
parameter. We see that the behavior of this integrated quantity is similar to the temperature of the peak of the heat capacity,
but the stabilization happens for $limit$=$10^{-4}$, $10^{-5}$, and $10^{-6}$.

Using this new variant for calculating the checking parameter we obtained results for $T_c$, $\gamma$ and $\beta$ from ten independent finite-size scaling simulations. In order to have a good ground for comparing both criteria, we repeated in this case exactly the same simulations that were performed in the previous section.

\begin{figure}[!hb]\centering
\begin{center}
 \includegraphics[width=.9\linewidth]{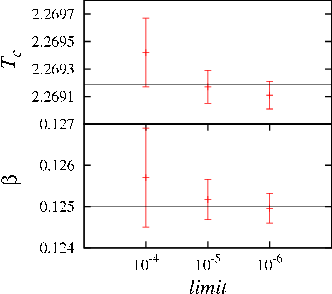}
\end{center}
\vspace{-.7cm}
\caption{(color online). Upper panel: Mean critical temperature from ten independent finite size scaling for three levels
of demand for the checking parameter, using the quantity $Q$ as reference. Lower panel: The exponent $\beta$ from the
same simulations. The straight lines in both panels refer to the exact values $T_c=2.2691853...$ and $\beta=0.125$}
\label{results_int}
\end{figure}

\begin{figure}[!h]\centering
\begin{center}
 \includegraphics[width=.9\linewidth]{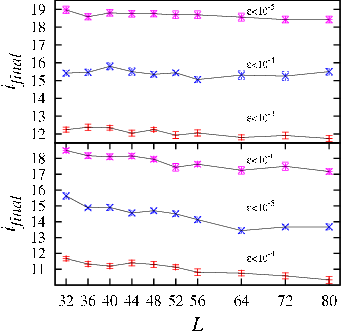}
\end{center}
\vspace{-.7cm}
\caption{(color online). Upper panel: Mean final order of the modification factor for three levels
of demand for the checking parameter, using $T_c(C)$ as reference. Lower panel: The same using
the quantity $Q$ for calculating the checking parameter}
\label{f_final}
\end{figure}

Top of Fig. \ref{results_int} displays the overall mean results for the critical temperature for
$limit$=$10^{-4}$, $10^{-5}$, and $10^{-6}$. Bottom shows the results for the
exponent $\beta$. The results for $\gamma$ are $1.7631(12)$, $1.7610(10)$, and $1.7605(10)$, respectively, again above the exact value, while the outcomes for the critical temperature and the exponent $\beta$ are in very good agreement with the exact data starting from $limit$=$10^{-4}$. In Fig. \ref{f_final} we show the mean final orders of the modification factor for each variant
of the checking parameter for three levels of demand. One can see that using $T_c(t)$ the orders are roughly size independent, but
if the quantity $Q(t)$ is adopted the orders of the final modification factor have a slight decrease with increasing lattice sizes.

\section{Results}
\vspace{.25cm}
\subsection{2D Ising model}
\vspace{.25cm}

Aware of the peculiarities of WLS, in order to test the reproductivity of the outcomes shown
above we carried out four more independent simulations using the two versions of checking
parameter. In Table \ref{table4} we show the results using the peak of the specific heat as
reference. The exponent $\gamma$ remains stubbornly above the exact value in all cases, but
regarding the critical temperature and the exponent $\beta$, we see that although the results
for $limit=10^{-3}$ are quite reasonable, the errorbars are in general larger then those of
the last two columns, which are roughly equivalent. The differences between the simulation
values and the exact values are in most cases less then the standard deviations $\sigma$ of
the simulations. Only the exponent $\beta$ in the fifth simulation is slightly beyond the
errorbars. Theses results give us confidence
to define $limit=10^{-4}$ as the ideal threshold for halting the simulations when the
checking parameter is defined by the temperature of the peak of the specific heat.

  \begin{table}[!]
  \centering
  \begin{tabular}{c c c c }
  \hline\hline
  $f_{13}$ & \hspace{.2cm} $\varepsilon<10^{-3}$ & \hspace{.2cm} $\varepsilon<10^{-4}$ & \hspace{.2cm} $\varepsilon<10^{-5}$ \\ [0.5ex]
  \hline
       & \hspace{0.3cm} $T_c$=2.2691853...      \\
  \hline
   a.  2.26917(15) & \hspace{-.2cm} 2.26925(18) & \hspace{-.2cm} 2.26916(10) & \hspace{.2cm} 2.26910(10) \\
   b.  2.26914(18) & \hspace{-.2cm} 2.26929(19) & \hspace{-.2cm} 2.26915(14) & \hspace{.2cm} 2.26911(14) \\
   c.  2.26915(11) & \hspace{-.2cm} 2.26909(16) & \hspace{-.2cm} 2.26913(10) & \hspace{.2cm} 2.26912(10) \\
   d.  2.26930(14) & \hspace{-.2cm} 2.26921(17) & \hspace{-.2cm} 2.26928(12) & \hspace{.2cm} 2.26927(10) \\
   e.  2.26925(15) & \hspace{-.2cm} 2.26916(16) & \hspace{-.2cm} 2.26928(15) & \hspace{.2cm} 2.26926(12) \\
   \hline
     &  \hspace{0.3cm}  $\gamma$=1.75       \\
   \hline
   a. 1.7609(12) & \hspace{-.2cm} 1.7600(11) & \hspace{-.2cm} 1.7605(10) & \hspace{.2cm} 1.7605(10) \\
   b. 1.7586(15) & \hspace{-.2cm} 1.7579(19) & \hspace{-.2cm} 1.7587(11) & \hspace{.2cm} 1.7589(10) \\
   c. 1.7597(15) & \hspace{-.2cm} 1.7601(18) & \hspace{-.2cm} 1.7584(11) & \hspace{.2cm} 1.7580(10) \\
   d. 1.7568(16) & \hspace{-.2cm} 1.7576(18) & \hspace{-.2cm} 1.7571(13) & \hspace{.2cm} 1.7573(10) \\
   e. 1.7590(11) & \hspace{-.2cm} 1.7590(15) & \hspace{-.2cm} 1.7577(10) & \hspace{.2cm} 1.7576(10) \\
  \hline
    & \hspace{0.3cm}   $\beta$=0.125      \\
   \hline
  a. 0.12520(63) & \hspace{-.2cm} 0.12541(72)  & \hspace{-.2cm} 0.12514(44) & \hspace{.2cm} 0.12491(32) \\
  b. 0.12526(72) & \hspace{-.2cm} 0.12540(79)  & \hspace{-.2cm} 0.12529(65) & \hspace{.2cm} 0.12523(60) \\
  c. 0.12560(56) & \hspace{-.2cm} 0.12561(56)  & \hspace{-.2cm} 0.12548(48) & \hspace{.2cm} 0.12516(45) \\
  d. 0.12572(74) & \hspace{-.2cm} 0.12528(84)  & \hspace{-.2cm} 0.12554(65) & \hspace{.2cm} 0.12547(52) \\
  e. 0.12559(52) & \hspace{-.2cm} 0.12559(56)  & \hspace{-.2cm} 0.12554(46) & \hspace{.2cm} 0.12554(34) \\
  \hline\hline
  \end{tabular}
  \caption{Five independent runs using the checking parameter defined by the peak of the specific heat. Each procedure
  is the result of ten finite-size scaling extrapolations for the critical temperature and the exponents $\gamma$ and
  $\beta$ for outputs at the end of $f_{13}$ and for decreasing $\varepsilon$, using the $80\%$ flatness criterion. }
  \label{table4}
  \end{table}

  \begin{table}[!ht]
  \centering
  \begin{tabular}{c c c c }
  \hline\hline
  $f_{13}$ & \hspace{-.3cm} $\varepsilon \leq10^{-4}$ & \hspace{-.3cm} $\varepsilon \leq10^{-5}$ & \hspace{.3cm} $\varepsilon \leq10^{-6}$ \\ [0.5ex]
  \hline
       & \hspace{0.3cm} $T_c$=2.2691853...      \\
  \hline
   a.  2.26917(15) & \hspace{-.3cm} 2.26942(25) & \hspace{-.5cm} 2.26917(12) & \hspace{.3cm} 2.26911(10) \\
   b.  2.26914(18) & \hspace{-.3cm} 2.26944(27) & \hspace{-.5cm} 2.26913(17) & \hspace{.3cm} 2.26911(14) \\
   c.  2.26915(11) & \hspace{-.3cm} 2.26921(16) & \hspace{-.5cm} 2.26918(11) & \hspace{.3cm} 2.26914(11) \\
   d.  2.26930(14) & \hspace{-.3cm} 2.26936(21) & \hspace{-.5cm} 2.26928(13) & \hspace{.3cm} 2.26927(11) \\
   e.  2.26925(15) & \hspace{-.3cm} 2.26959(20) & \hspace{-.5cm} 2.26926(15) & \hspace{.3cm} 2.26925(13) \\
   \hline
     &  \hspace{0.3cm}  $\gamma$=1.75       \\
   \hline
   a.  1.7609(12)  & \hspace{-.3cm} 1.7631(12) & \hspace{-.5cm}  1.7610(10) & \hspace{.3cm} 1.7605(10) \\
   b.  1.7586(15)  & \hspace{-.3cm} 1.7590(21) & \hspace{-.5cm}  1.7582(12) & \hspace{.3cm} 1.7593(10) \\
   c.  1.7597(15)  & \hspace{-.3cm} 1.7615(14) & \hspace{-.5cm}  1.7590(11) & \hspace{.3cm} 1.7587(11) \\
   d.  1.7568(16)  & \hspace{-.3cm} 1.7568(16) & \hspace{-.5cm}  1.7568(16) & \hspace{.3cm} 1.7571(10) \\
   e.  1.7590(11)  & \hspace{-.3cm} 1.7588(11) & \hspace{-.5cm}  1.7589(11) & \hspace{.3cm} 1.7577(10) \\
  \hline
    & \hspace{0.3cm}   $\beta$=0.125      \\
   \hline
   a.  0.12520(63) & \hspace{-.3cm} 0.1257(12)  & \hspace{-.5cm}  0.12517(49) & \hspace{.3cm} 0.12496(36) \\
   b.  0.12526(72) & \hspace{-.3cm} 0.12560(97) & \hspace{-.5cm}  0.12525(68) & \hspace{.3cm} 0.12520(60) \\
   c.  0.12559(56) & \hspace{-.3cm} 0.12588(79) & \hspace{-.5cm}  0.12550(52) & \hspace{.3cm} 0.12522(45) \\
   d.  0.12572(74) & \hspace{-.3cm} 0.12619(85) & \hspace{-.5cm}  0.12565(67) & \hspace{.3cm} 0.12543(57) \\
   e.  0.12559(52) & \hspace{-.3cm} 0.12655(71) & \hspace{-.5cm}  0.12557(50) & \hspace{.3cm} 0.12549(40) \\
  \hline\hline
  \end{tabular}
  \caption{Five independent runs using the checking parameter defined by the heat transfer. Each procedure
  is the result of ten finite-size scaling extrapolations for the critical temperature and the exponents $\gamma$ and
  $\beta$ for outputs at the end of $f_{13}$ and for decreasing $\varepsilon$, using the $80\%$ flatness criterion. }
  \label{table5}
  \end{table}

Table \ref{table5} shows the results for the same independent finite-size scaling simulations using the heat transfer to
define the checking parameter for halting the process. We see that the overall results for $limit=10^{-4}$ are inaccurate.
One can observe that the results using $limit=10^{-5}$ and $limit=10^{-6}$ are equivalent. We conclude therefore that
if the second criterion is being used the ideal value to stop the simulations should be $limit=10^{-5}$.

  If adaptive windows \cite{adaptive} are being used in the simulations, both checking parameters can be applied. For
  calculating the peak of the specific heat or the heat transfer during the simulations one should just virtually joint
  the current windows. As demonstrated in Ref. \cite{adaptive} the use of fixed overlapping windows is not advisable in
  many models, but in case that it is being adopted, the second checking parameter may still be applied. Instead of
  taking the canonical averages defined in Eq. (\ref{mean}) over all the energy spectrum, one should carry out the
  summations over the energy levels of each window. Naturally the resulting calculation looses the meaning of
  specific heat and the integral of Eq. (\ref{heat}) either can be interpreted as the heat transfer, but this integrated value
  can nonetheless be used to define the checking parameter. In order to demonstrate the validity of such a procedure,
  we divided the spectrum of the $L=32$ Ising model lattice in four windows and carried out $N=24$ independent
  simulations for each of them using $limit=10^{-5}$ for halting the simulations. The resulting mean final orders of
  the modification factor were 14.32(21), 14.65(18), 14.43(20), and 14.71(19), respectively, which are compatible with the
  result of Fig. \ref{f_final} using the full spectrum, although slightly bellow.

  Such large repetitious handling of data for obtaining all these canonical averages
  and  finite-size extrapolations were possible only by using shell scripting
  \cite{AdvBashScr,BGB2008,Robbins2005,Neves2008,Jargas2008-Shell}. This is an exceptional
  tool for those who work with simulations.

\subsection{3D Ising model}
\begin{figure}[!h]\centering
\begin{center}
 \includegraphics[width=.9\linewidth]{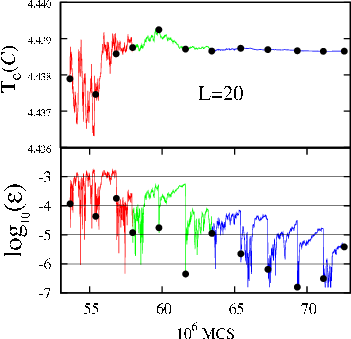}
\end{center}
\vspace{-.7cm}
\caption{(color online). Upper panel: Evolution of the temperature of the maximum of the specific heat during the WLS of
the 3D Ising model, beginning from $f_{8}$ for a single run. The dots show where the modification factor was updated. Lower panel: Evolution
of the logarithm of the checking parameter $\varepsilon$ during the same simulation.}
\label{3d_ising}
\end{figure}

\vspace{.25cm}
In order to illustrate the applicability of this technique to a three-dimensional system, we performed WLS for
the 3D Ising model on a $20\times20\times20$ lattice. The behavior of the canonical averages is similar to the
2D Ising model and shows that our method for halting the simulations can be applied to a three-dimensional
system as well. Top of Fig. \ref{3d_ising} shows the evolution of the temperature of the peak of the specific
heat beginning from $f_8$, and bottom displays the logarithm of the checking parameter. In this case the
simulations would stop at the end of $f_{10}$, $f_{13}$, and $f_{18}$ for $limit=10^{-3}$, $10^{-4}$,
and $10^{-5}$, respectively.

\subsection{Homopolymer}
\vspace{.25cm}
As a new application of the criterion, we consider a homopolymer consisting of $N$ monomers which may assume any self avoiding
walk (SAW) configuration on a two-dimensional lattice \cite{bjp,cpc}. Assuming that the polymer is in a bad solvent, there is an effective monomer-monomer attraction in addition to the self-avoidance constraint representing excluded volume. For every pair of
non-bonded nearest-neighbor monomers the energy of the polymer is reduced by $\epsilon$. The Hamiltonian for the model
can be written as

\begin{equation}\mathcal H=-\epsilon\sum_{<i,j>}{\sigma_i\sigma_j},
\end{equation}
where $\sigma=1(0)$ if the site $i$ is occupied(vacant), and the sum is over nearest-neighbor pairs\cite{dickman}. (The
sum is understood to exclude pairs of bonded segments along the chain.) We used the so-called reptation or
\textit{``slithering snake''} move which consists of randomly adding a monomer to one end of the chain and removing
a monomer from the other end, maintaining the size of the polymer constant. (Although reptation is not suitable for sampling
the most compact configurations, this does not affect the conclusions presented here.) We define one Monte Carlo step
as $N$ attempted moves.

\begin{figure}[!ht]\centering
\begin{center}
 \includegraphics[width=.9\linewidth]{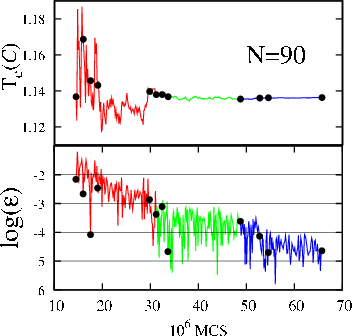}
\end{center}
\caption{(color online). Upper panel: Evolution of the temperature of the maximum of the specific heat of the self-avoiding homopolymer of $L=90$ during the WLS, beginning from $f_{10}$ for a single run. The dots show where the modification factor was updated. Lower panel: Evolution of the logarithm of the checking parameter $\varepsilon$ during the same simulation.}
\label{check_pol}
\end{figure}

\begin{figure}[!ht]\centering
\begin{center}
 \includegraphics[width=.7\linewidth, angle=-90]{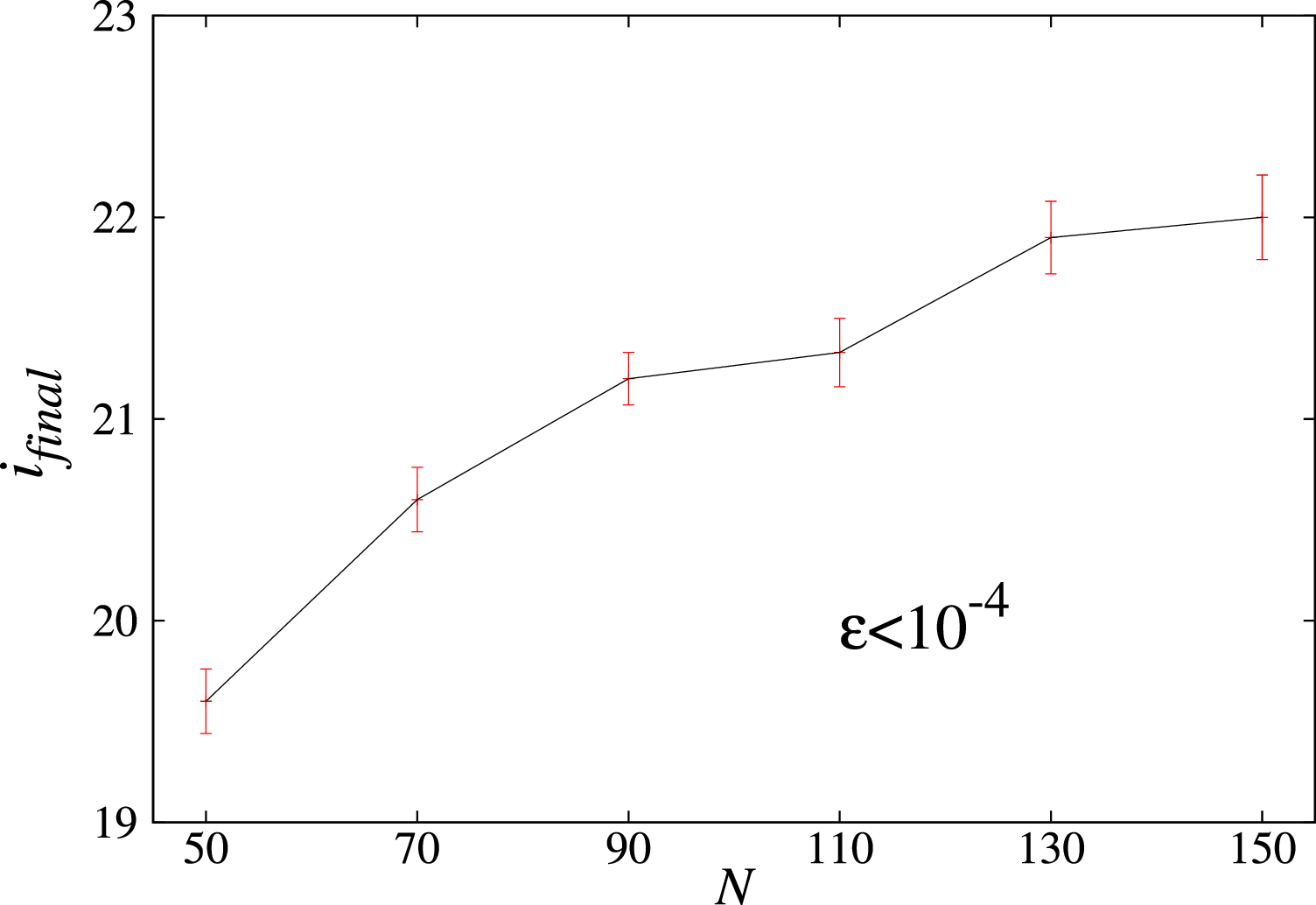}
\end{center}
\caption{(color online). Mean order of the final modification factor for each simulated size of homopolymers using
  the $80\%$ flatness criterion.}
\label{f_final_pol}
\end{figure}

We performed simulations for polymers of sizes $N = 50, 70,..., 150$ taking $10$ independent runs for each size
and finishing the process when the condition
\begin{equation}\label{eps}
 |T_c(t)-T_c(0)|<10^{-4}
\end{equation}
was satisfied in the course of the whole simulation of a  modification factor, where again, $T_c(0)$ is the last temperature
of the peak of the specific heat in the previous modification factor.

In Fig.\ref{check_pol} we show the evolution of the temperature of the peak of the specific heat of a polymer of $N=90$,
beginning from $f_{10}$. In the lower panel the logarithm of the checking parameter $\varepsilon$ of the same simulation is displayed. The simulations were started from constructed ground state configurations\cite{ground}.

In Fig \ref{f_final_pol} we show the mean order of the final modification factor for each polymer size.
One can see that unlike the two-dimensional Ising model, in this case the order of the final
modification factor increases with increasing polymer sizes. Adopting the proposed criterion for halting
the simulations ensures that each particular run proceeds up to the real stabilization of the results.

Often one faces difficulties in sampling conformations with lowest energies and reaching the flatness criterion. Many
ingenious protocols of evolution of polymers have been proposed, e. g. Ref. \cite{wust}. The use of the criterion for
halting the simulations developed in this work can enhance accuracy and save a lot of CPU time.

\maketitle

\section{Conclusions}
\vspace{.25cm}
We proposed a criterion to finish the simulations of the Wang-Landau sampling. Instead of determining a final
modification factor $f_{final}$ for all simulations and every lattice sizes, the behavior of the temperature
of the peak of the specific heat or the heat transfer per unit are checked during the simulations and the process is halted
when these values vary bellow a given limit during the whole simulation of a modification factor. As a result,
different runs stop at different final modification factors. Our results define that the ideal $limit$ should be
$1^{-4}$ if the peak of the specific heat is used and $10^{-5}$ if one adopts the heat transfer per unit. We demonstrated
that reliable and accurate results are obtained only by taking averages over manifold simulations.  We applied
this technique to the two-dimensional Ising model and a homopolymer and found that for the Ising model the mean
final order of the modification factors are roughly the same for all lattice sizes, but for the homopolymer the
final order of the modification factor increases with increasing polymer sizes. We also presented a brief
application of the criterion to the three-dimensional Ising model.

Finally it should be pointed out that Wang-Landau variants that do not require the use of the modification factor,
e. g. Ref. \cite{belardinelli3,eisenbach2} may benefit of the ideas proposed in this work by defining a suitable
number of MCS, after which the checking parameter would be calculated. If $\varepsilon$ remains bellow a
predefined $limit$ for, say, ten consecutive checks the simulation could be halted.

\section{Acknowledgment}

This work was supported by FUNAPE-UFG. We acknowledge the computer resources provided by LCC-UFG and thank
Salviano de Ara{\'u}jo Le{\~a}o for his helpful and substantial advice and support with the
computations.

\end{document}